\documentclass[letterpaper, 10 pt, conference]{ieeeconf}  

\IEEEoverridecommandlockouts                              
\usepackage[utf8]{inputenc}
\usepackage[T1]{fontenc}
\usepackage{xcolor}

\usepackage[utf8]{inputenc}
\usepackage[T1]{fontenc}

\usepackage{amsmath,amsfonts}
\newcommand*{\transpose}{^{\mkern-1.5mu\mathsf{T}}}  
\usepackage{siunitx}
\DeclareMathOperator{\diag}{diag}

\usepackage{bm}
\usepackage{algpseudocode}
\usepackage{algorithm}

\usepackage[caption=false]{subfig}
\usepackage{tikz}
\usetikzlibrary{arrows.meta, automata, positioning}

\usepackage{array}
\usepackage{textcomp}
\usepackage{stfloats}
\usepackage{url}
\usepackage{verbatim}
\usepackage{graphicx}

\usepackage{booktabs}
\usepackage{tabularx}
\usepackage{adjustbox}

\usepackage{siunitx} 
\usepackage{gensymb}
\usepackage{cite}
\usepackage{balance}

\usepackage{dirtytalk} 
\MakeRobust{\say} 

\makeatletter
\let\NAT@parse\undefined
\makeatother

\usepackage[colorlinks,linkcolor=blue,anchorcolor=black,citecolor=blue,urlcolor=black,hyperfootnotes=true]{hyperref}   
\usepackage[all]{hypcap}

\title{\LARGE
Energy-based Regularization for Learning Residual Dynamics\\in Neural MPC for Omnidirectional Aerial Robots
}

\author{Johannes Kübel$^{1*}$, Henrik Krauss$^{2}$, Jinjie Li$^{1}$, and Moju Zhao$^{1}$
    \thanks{$^*$Corresponding author.}
    \thanks{$^{1}$DRAGON Lab, Department of Mechanical Engineering, The University of Tokyo, Tokyo 113-8656, Japan
            {\tt\small \{johannes-kubel, jinjie-li, chou\}@dragon.t.u-tokyo.ac.jp}}%
    \thanks{$^{2}$Department of Advanced Interdisciplinary
Studies, The University of Tokyo, Tokyo, 153-8904, Japan {\tt\small henrik1.krauss@gmail.com}}
    \thanks{This work was supported by Japan's Ministry of Education, Culture, Sports, Science and Technology (MEXT) under grant no. 241237 and by JST ASPIRE (JPMJAP2521).}
}

\begin{document}
\bstctlcite{IEEEtran:BSTcontrol}  

\maketitle
\thispagestyle{empty}
\pagestyle{empty}

\begin{abstract}
Data-driven Model Predictive Control (MPC) has lately been a core research subject in the field of control theory. The combination of an optimal control framework with deep learning paradigms opens up the possibility to accurately tracking control tasks without the need for complex analytical models. However, the system dynamics are often nuanced and the neural model lacks the potential to understand physical properties such as inertia and conservation of energy. In this work, we propose a novel \textit{energy-based regularization} loss function which is applied to the training of a neural model that learns the residual dynamics of an omnidirectional aerial robot. Our energy-based regularization encourages the neural network to learn stabilizing control corrections. Without the regularization the network misses physical context. The learned residual dynamics are then integrated into the MPC framework. The positional mean absolute error (MAE) are shown to be improved in three real-world experiments by 23\% compared to the analytical MPC. We also compare our method to a standard neural MPC implementation without regularization and primarily achieve an increased flight stability implicitly due to smoothening out the acceleration of the robot and thus up to 15\% lower MAE. A complimentary video is available at:~\url{https://youtu.be/E3XG979NXHk}.
We share our code under:~\url{https://github.com/johanneskbl/jsk_aerial_robot/tree/develop/neural_MPC}.
\end{abstract}

\section{Introduction} \label{sec:intro}
It is crucial to design control systems that accurately compute control signals that automatically regulate the state of systems in many applications. Nowadays, the applications continue to become more complex such as in aerial robots, autonomous vehicles, or logistic units. Specifically omnidirectional aerial robots are gaining increasing attention both from academia and industry~\cite{ollero_past_2022, nishio_design_2024, li_six_dof_2025}. Unlike traditional underactuated aerial robots that lack the ability to uncouple their attitude and position control, omnidirectional aerial robots can fully actuate their pose but are significantly more challenging to control accurately~\cite{ryll_6d_2019, allenspach_design_2020, yigit_dynamic_2023, zhong_prototype_2024}. Previous works especially focused on tiltable-quadrotors as a subclass of omnidirectional aerial robots.

One popular control method for tiltable-quadrotors is Model Predictive Control which uses a model-driven approach to compute an optimal control input~\cite{li_servo_2024}. The main advantages of MPC lie in the problem formulation which can be designed to optimize a system-specific control goal even given state and input constraints.
At its core, MPC relies heavily on the dynamical model of the system to optimize the system behavior over the control input. When MPC is used for complex systems, such as aerial robots, where the true dynamical system, including disturbances, is difficult to model, the accuracy in trajectory tracking often leaves room for improvement and requires additional techniques~\cite{li_servo_2024}. The sensitivity of MPC to the underlying model is a central disadvantage of the methodology.

\begin{figure}
    \centerline{\includegraphics[width=0.48\textwidth]{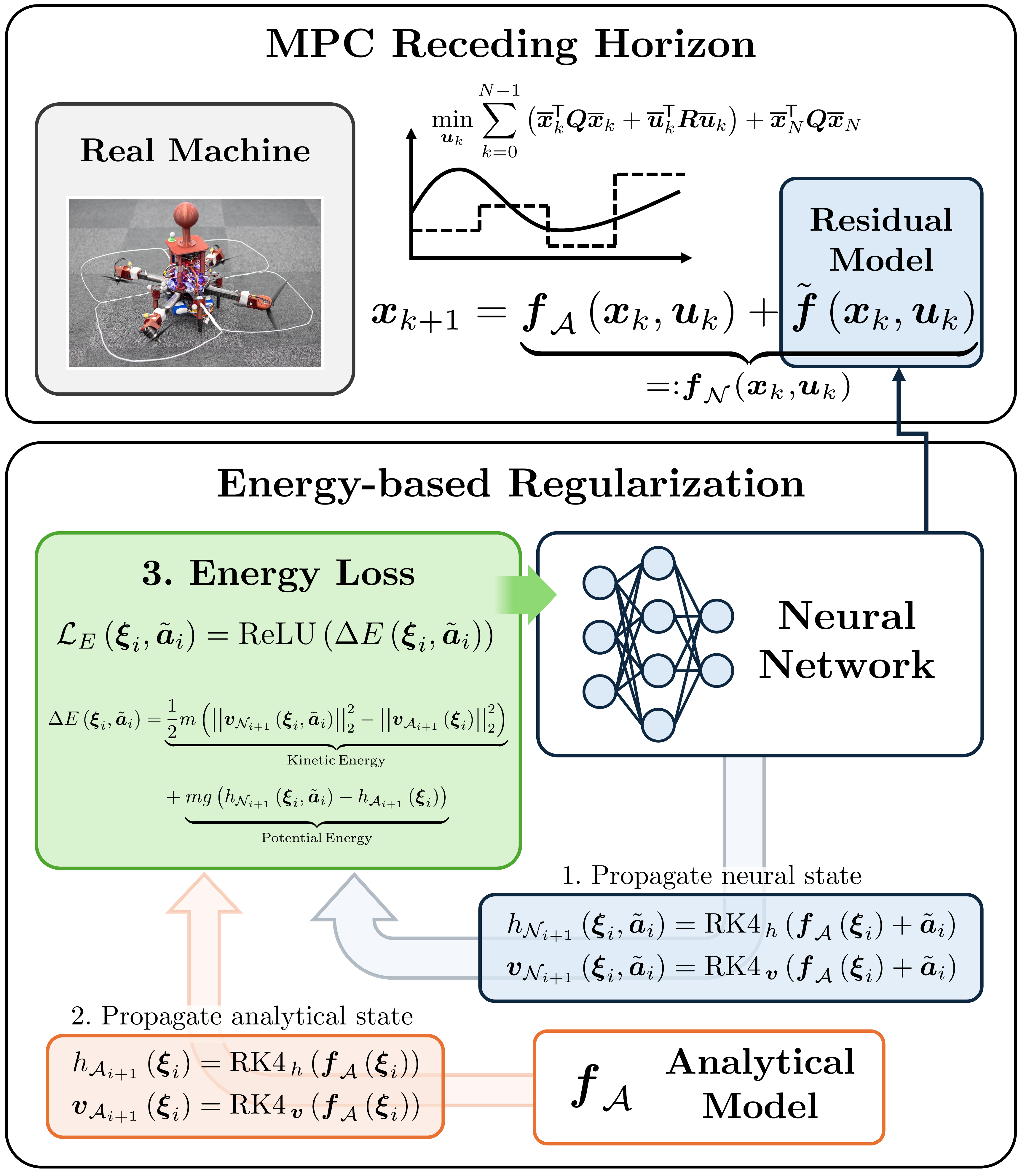}}
    \caption{Schematic overview of our proposed \textit{Energy-based Regularization for Neural MPC}. The upper Neural MPC framework containing an analytical model and a neural network controls the aerial robot. The lower training framework shows how our energy-based regularization technique enhances the learning of the network. We first capture the energy of the predicted state following the analytical and neural-enhanced model. Then, the difference between the energy states builds up the energy loss that encourages the network to output stabilizing corrections in the dynamics due to the added physical context.}
    \label{fig:overview_loss}
\end{figure}

Accurately modeling aerial robots with classical mechanics proves to be a challenge since the dynamics for the airflow generated by the rotors is highly turbulent and introduces vortices that depend on the air density, temperature, and other external influences. Especially the interaction with the surrounding environment through the turbulent airflow factors in significantly to the forces acting on the system. A key example of these forces is the concept of \textit{ground effect} where the robot experiences a force directed away from the ground, generally accelerating the robot upwards against gravity. Furthermore, the interaction with the ground is destabilizing the system since the disturbance fluctuates in its direction and magnitude depending on the distance to the ground, the angle of attack, and airspeed~\cite{panerati2021learning}. Machine learning techniques have been shown to be able to capture complex nonlinear dependencies in a system. Therefore, recent research has aimed to bridge the gap between classical optimal control and deep learning inside the MPC framework~\cite{brunke2022safe, li2023nonlinear, zhang2025proxfly, salzmann2023real}.
Even though many advances have been made in this field, the integration of the neural model into the MPC framework still is a cutting-edge topic. Specifically, flight stability is often traded for point-wise accuracy requiring additional considerations regarding the effects of the network output on the system. General approaches simply use an $\mathcal{L}_{2}$ norm loss function to train the neural network but our method proposes an additional \textit{energy-based regularization} loss that encodes the physical effect of the network output on the energy of the system. Based on the network output the state is propagated by using both the analytical and neural-enhanced model. Then the energy-based loss measures the change of kinetic and potential energy in the system compared on both models. By penalizing if the network output increased the energy of the system, we aim to train the network to prefer outputs that induce lower energy and therefore lower acceleration of the system. Ultimately the introduction of the energy-based regularization then stabilizes the system. Leveraging the underlying assumption that, if the system's energy is lowered, the system is less sensitive to imprecise model predictions, the flight behavior can be stabilized and more robust.

In this work, we apply the idea of data-driven MPC to omnidirectional aerial robots while making use of the novel regularization. Specifically, we
\begin{enumerate}
    \item build a framework for neural MPC applied to general aerial robots, 
    \item design an \textit{energy-based regularization} technique to train a neural model for the MPC controller which is subject to the physical context, and
    \item validate our proposed neural MPC in a set of real-world experiments against two benchmark approaches showing that the network is able to learn from recorded data to compensate model error and other disturbances while achieving higher accuracy and stabilized flight.
\end{enumerate}

To leverage the use of our proposed energy-based neural MPC, we apply our method to an omnidirectional aerial robot which due to its design is providing more actuation dimensions than degrees of freedom proving to be a challenging system to control. For real-world experiments we build a similar system to the tiltable-quadrotor presented in~\cite{li_servo_2024}. The tiltable-quadrotor is designed with servo motors at the rotors that allow for vectorization of the thrust.

\section{Related Work} \label{sec:related_work}
The past research into data-driven MPC tackles the underlying problem from various directions applied to different systems. Torrente et al.~\cite{torrente2021data} were among the first to apply the concept to quadrotors using Gaussian Processes (GP) to model aerodynamic drag effects that the nominal model cannot capture. While GPs provide principled uncertainty estimates, their computational complexity scales cubically with training data, which severely limits their applicability to embedded real-time systems. Salzmann et al.~\cite{salzmann2023real} address the scalability problem by using local approximations of fully connected networks to maintain real-time feasibility even for large networks. We use their Real-Time Neural MPC (RTNMPC) method as our baseline in section~\ref{sec:experiments}. Another approach is to improve the analytical model itself to avoid any learning methods which is implemented in~\cite{li_servo_2024} and which we use for our analytical MPC benchmark. Recent work explores reinforcement learning as an alternative to MPC for tiltable-quadrotors~\cite{zhang2026learning}, achieving robustness through sim-to-real transfer, but fundamentally abandoning the model-based structure that allows interpretable constraint handling and deterministic optimization. Learning residual dynamics represented by a fully-connected neural network to augment a nominal analytical model in the MPC framework is also realized in~\cite{jiang2022neural, li2023nonlinear}.

Importantly, all of the above methods using a residual network train it with an $\mathcal{L}_2$ loss on the predicted state and label. Other works show the need for regularization techniques and propose approaches based on machine learning paradigms such as stochastic regularization and Variational Auto Encoders (VAE)~\cite{osaka2025deterministic,yin2021augmenting,takeishi2023deep}. However, all of these supervised learning objectives are agnostic to the physical implications of the predicted output and provide no incentive for the trained network to avoid nonphysical or even destabilizing corrections. The idea of embedding physical knowledge into the training of neural network dynamics models has been investigated by Physics Informed Neural Networks (PINNs)~\cite{raissi2017physics}. Our energy-based regularization loss has a similar motivation and penalizes the kinetic and potential energy difference between the state predicted by the neural-enhanced model and the analytical model. To the best of our knowledge, no prior work has applied a similar technique to the learning of residual dynamics used in MPC.

\section{Analytical Model} \label{sec:analytical_model}
\begin{figure}
    \centerline{\includegraphics[width=0.465\textwidth]{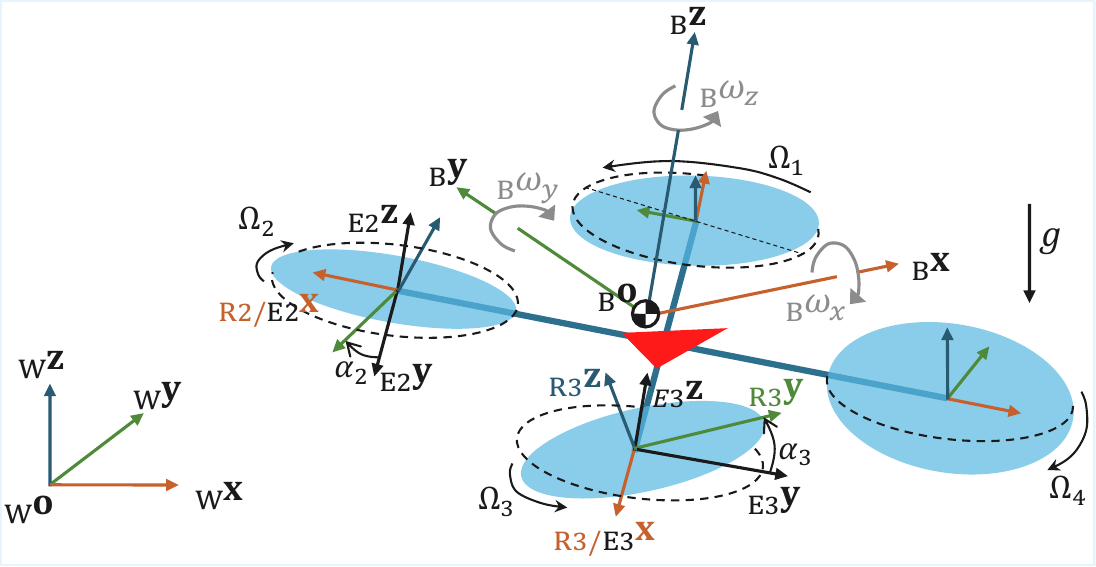}}
    \caption{Schematic overview of a tiltable-quadrotor with four coordinate frames: \say{World} $\left\{\mathcal{W}\right\}$ in ENU (X East, Y North, Z Up), \say{body} $\left\{\mathcal{B}\right\}$ in FLU (X Forward, Y Left, Z Up), \say{end-of-arm} $\left\{\mathcal{E}_r\right\}$, and \say{rotor} $\left\{\mathcal{R}_r\right\}$ for each rotor $r$.}
    \label{fig:coordinate_systems}
\end{figure}

As preliminaries, we denote scalars with $x, X \in \mathbb{R}$, vectors with bold lowercase $\boldsymbol{x} \in \mathbb{R}^s$, and matrices with bold uppercase $\boldsymbol{X} \in \mathbb{R}^{s \times m}$. Unit axes of coordinate frames are written in bold roman $\mathbf{x}, \mathbf{y}, \mathbf{z} \in \mathbb{R}^3$. For coordinate transforms, a vector in the body frame $\left\{\mathcal{B}\right\}$ is written as\,$_\mathrm{B}\boldsymbol{p}$ and 3D rotations from the rotor frame $\left\{\mathcal{R}\right\}$ to the world frame $\left\{\mathcal{W}\right\}$ are represented by the quaternion\,$_\mathrm{W}\boldsymbol{q}$.
The inertial world frame is defined by 
\( \left\{\mathcal{W}\right\} = \left\{\,_\mathrm{W}\mathbf{o}, \,_\mathrm{W}\mathbf{x}, \,_\mathrm{W}\mathbf{y}, \,_\mathrm{W}\mathbf{z} \right\} \) 
with its\,$_\mathrm{W}\mathbf{z}$ axis opposite to the direction of gravity. For readability, we write any variable in the world frame as $\boldsymbol{p}$ instead of $_\mathrm{W}\boldsymbol{p}$.
The schematic model of the tiltable-quadrotor can be seen in Figure~\ref{fig:coordinate_systems} including all coordinate systems used. The origin of the body frame $\left\{\mathcal{B}\right\}$ is located at the quadrotor's CoG with its\,$_\mathrm{B}\mathbf{x}$-axis to the front and its\,$_\mathrm{B}\mathbf{z}$-axis upwards. The coordinate systems $\left\{\mathcal{E}_r\right\}$ describe the end of each arm $r$ and its direction with\,$_{\mathrm{E}_r}\mathbf{x}$-axis pointing outwards. The transformations\,$_\mathrm{B}\boldsymbol{R}\,_{\mathrm{E}_r}$ can be calculated with the quadrotor's physical properties.
The rotor frames $\left\{\mathcal{R}_r\right\}$ coincide with $\left\{\mathcal{E}_r\right\}$ but are rotated by the servo angles $\alpha_r$ which also control the angle of attack for each thrust vector. Therefore, the coordinate transformation for rotor $r$ is defined by the standard rotation matrix around\,$_{\mathrm{E}_r}\mathbf{x}$.

The derivation of the analytical model is based on the theory of classical mechanics. First, we define the state
\begin{equation}
    \boldsymbol{x} = \left[\boldsymbol{p}\transpose, \, \boldsymbol{v}\transpose,\,  \boldsymbol{q}\transpose, \,_\mathrm{B}\boldsymbol{\omega}\transpose, \, \boldsymbol{\alpha}\transpose\right]\transpose
\end{equation}
consisting of the position $p$ of the center of gravity (CoG) and its velocity $v$ as well as the quadrotor's attitude defined by the quaternion $q$ and the angular velocity by\,$_\mathrm{B}\omega$. 
Note that the angular velocity is given in the body frame $\{\mathcal{B}\}$ to allow for a simpler expression of the angular acceleration dynamics.

\subsection{Rotor Dynamics}
For aerial robots the main force contribution in its dynamics is the thrust generated by the rotors. In our tiltable-quadrotor the thrust at each rotor $r$ is modeled by the widely used quadratic fit
\begin{equation}\label{eq:thrust_model}
    f_r = k_t \Omega_r^2, ~~~~~ \tau_r = k_q \Omega_r^2,
\end{equation}
where $\Omega_r$ is the rotors rotation speed and the parameters $k_t$ and $k_q$ are rotor coefficients determined through parameter identification.
We use $d_r$ to describe the rotation direction and\,$_\mathrm{B}\boldsymbol{p}'_r$ for the rotors position in $\left\{\mathcal{B}\right\}$.

\subsection{Servo Dynamics}
To achieve omnidirectional flight we use servo motors at the end of each arm to tilt the rotors by an angle $\alpha_r$ around the\,$_{\mathrm{R}_r}\mathbf{x}$-axis. In~\cite{li_servo_2024} the authors included the servo angle as a state $\alpha_{s_r}$ in the analytical model for the MPC formulation. The servo angle dynamic is assumed to follow the first-order model
\begin{equation}
    \dot{\boldsymbol{\alpha}}_s = \frac{1}{T_{\mathrm{servo}}} (\boldsymbol{\alpha}_c - \boldsymbol{\alpha}_s),
\end{equation}
where $\boldsymbol{\alpha}_c$ is the control command for the servo motors and $T_{\mathrm{servo}} = \SI{0.048}{s}$ is the servo motors' time constant. We use a similar approach and refer to~\cite{li_servo_2024} for a detailed analysis of including the servo angle state in the model as well as the allocation of actuator references.

\subsection{Wrench from Rotor to Body Frame}
The wrench induced to the system by the rotor $r$ expressed in $\left\{\mathcal{R}_r\right\}$ follows
\begin{subequations}
    \begin{align}
        _{\mathrm{R}_r}\boldsymbol{f}_r &= [0, ~ 0, ~ f_r]\transpose, ~~~~~ f_r \in \left[f_{\mathrm{min}}, f_{\mathrm{max}}\right], \\
        _{\mathrm{R}_r}\boldsymbol{\tau}_r &= \left[0, 0, -d_r f_r \frac{k_q}{k_t}\right] \transpose.
    \end{align}
\end{subequations}

Transforming the wrench into $\left\{\mathcal{B}\right\}$ results in
\begin{subequations}
\label{eq:resultant_wrench}
    \begin{align}
        {_\mathrm{B}\boldsymbol{f}_u} = \sum_{r=1}^{4}&{\,_\mathrm{B}\boldsymbol{R}\,_{\mathrm{E}_r}} {\,_{\mathrm{E}_r}\boldsymbol{R}\,_{\mathrm{R}_r}} {\,_{\mathrm{R}_r}\boldsymbol{f}_r}, \label{eq:fu} \\
            {_\mathrm{B}\boldsymbol{\tau}_u} = \sum_{r=1}^{4}&{\,_\mathrm{B}\boldsymbol{R}\,_{\mathrm{E}_r}} {\,_{\mathrm{E}_r}\boldsymbol{R}\,_{\mathrm{R}_r}}{\,_{\mathrm{R}_r}\boldsymbol{\tau}_r} \label{eq:tau_u} \\
            & +\,_\mathrm{B}\boldsymbol{p}'_{i} \times{{_\mathrm{B}\boldsymbol{R}\,_{\mathrm{E}_r}} {\,_{\mathrm{E}_r}\boldsymbol{R}\,_{\mathrm{R}_r}}{\,_{\mathrm{R}_r}\boldsymbol{f}_r}}. \nonumber
    \end{align}
\end{subequations}

\subsection{Full Dynamics}
Using the wrench expressed in $\left\{\mathcal{B}\right\}$ we can derive the dynamic of the state with fundamentals from rigid-body mechanics. The time-derivative of the state is given by
\begin{subequations}\label{eq:full_dynamics}
    \begin{align}
        \dot{\boldsymbol{p}} &= \boldsymbol{v}, \\
        \dot{\boldsymbol{v}} &= \frac{1}{m} \left(\,_\mathrm{W}\boldsymbol{R}\,_\mathrm{B}(\boldsymbol{q}) {\,_\mathrm{B}\boldsymbol{f}_u} \right) - \boldsymbol{g}, \label{eq:linear_analytical_dynamics}\\
        \dot{\boldsymbol{q}} &= \frac{1}{2} \boldsymbol{q} ~ \circ ~ \mathcal{H}(_\mathrm{B}\boldsymbol{\omega}), \\
        {_\mathrm{B}\dot{\boldsymbol{\omega}}} &=\boldsymbol{{I}}^{-1} \left(-{_\mathrm{B}\boldsymbol{\omega}} \times\left({\boldsymbol{{I}}}\,{_\mathrm{B}\boldsymbol{\omega}}\right)+{_\mathrm{B}\boldsymbol{\tau}_u} \right), 
    \end{align}
\end{subequations}
where the physical parameters are the mass $m$, inertia matrix $\boldsymbol{I}=\diag ({I}_{xx}, {I}_{yy}, {I}_{zz})$ and gravity ${\boldsymbol{g}}=[0,0,g]\transpose$. For the quaternion dynamics we use $\circ$ to denote the quaternion multiplication and $\mathcal{H}(\cdot)$ as an operator to extend a 3D vector $\mathcal{H}(\cdot) := [\boldsymbol{0}, \cdot]\transpose$.

\section{Methodology} \label{sec:e_nmpc}
\subsection{Analytical MPC}
Model Predictive Control is an optimal control method to solve Nonlinear Programming (NLP) problems that uses a mathematical model of the underlying system to simulate its behavior over time. MPC iteratively predicts the state trajectory with varying control inputs. By defining a cost function we can evaluate the evolution of the state and input and optimize the system's response over the control input.

A classical optimal control problem that is discretized over time can be formulated with
\begin{subequations} \label{eq:analytical_problem}
    \begin{align}
        \underset{\boldsymbol{u}_k}{\rm min} \quad &\sum\limits_{k=0}^{N-1}\left(\overline{\boldsymbol{x}}\transpose_k \boldsymbol{Q} \overline{\boldsymbol{x}}_k + \overline{\boldsymbol{u}}\transpose_k \boldsymbol{R} \overline{\boldsymbol{u}}_k\right) + \overline{\boldsymbol{x}}\transpose_N \boldsymbol{Q}\overline{\boldsymbol{x}}_N, \label{eq:cost} \\
        {\rm s.t.} \quad
        &\boldsymbol{x}_{k+1} = \boldsymbol{f}\left(\boldsymbol{x}_{k}, \boldsymbol{u}_{k}\right), \quad ~~~ k=0, \ldots, N-1, \label{eq:dyn_constraint} \\
        &\boldsymbol{x}_0 =\hat{\boldsymbol{x}}, \label{eq:init_constraint} \\
        &\boldsymbol{x} \in \mathcal{X}, \boldsymbol{u} \in \mathcal{U}, \label{eq:space_constraint}
    \end{align}
\end{subequations}
where~\eqref{eq:cost} defines, without loss of generality, a minimization of the cost function over the discretized control inputs $\boldsymbol{u}_k$ for all nodes $k\in[0,N-1]$ within the time horizon $N$. For tracking a reference trajectory $\boldsymbol{x}_{\mathrm{ref}}$ and $\boldsymbol{u}_{\mathrm{ref}}$ we define the tracking error
\begin{align}
        \overline{\boldsymbol{x}} = \boldsymbol{x}_{\mathrm{ref}} - \boldsymbol{x}, \,\,\,\, \overline{\boldsymbol{u}} = \boldsymbol{u}_{\mathrm{ref}} - \boldsymbol{u}.
\end{align}
Note we allocate the reference as in~\cite{li_servo_2024}. The optimization problem is constrained by the dynamical model $\boldsymbol{f}\left(\boldsymbol{x}_{k}, \boldsymbol{u}_{k}\right)$ in~\eqref{eq:dyn_constraint} which, for the standard MPC formulation, consists of the analytical model defined in section~\ref{sec:analytical_model}. The problem formulation is stated here as an initial value problem given the initial condition $\hat{\boldsymbol{x}}$. The formulation also seamlessly allows for state and input constraints $\mathcal{X}$, $\mathcal{U}$ in~\eqref{eq:space_constraint}, in our case simple box constraints.

Sequential Quadratic Programming (SQP) can be used to solve the optimization problem~\eqref{eq:analytical_problem} iteratively resulting in an optimized control sequence $\boldsymbol{u}^* = \{\boldsymbol{u}^*_0, \ldots, \boldsymbol{u}^*_{N-1}\}$. SQP with direct multiple shooting is a popular approach to enable the MPC scheme to be runtime viable by segmenting the optimization problem into parts and solving them in parallel given additional continuity constraints. The control inputs are optimized over the entire horizon in $T_{\mathrm{step}}$ time intervals and at each sampling time $T_{\mathrm{samp}}$ the first input from the optimized control sequence $\boldsymbol{u}^*$ is applied to the system.

\subsection{Neural MPC}
A central concept in MPC is propagating the state $\boldsymbol{x}_k$ to the next time step $k+1$ given a control input $\boldsymbol{u}_k$ through the model $\boldsymbol{f}(\boldsymbol{x}_k,\boldsymbol{u}_k)$. MPC faces the underlying challenge of depending on the accuracy of the dynamical model $\boldsymbol{f}$ which often only captures the main components in the real dynamics and ignores effects that are difficult to model analytically.

A common approach to compensate model errors is to learn from real-world data~\cite{salzmann2023real, abu2022deep, lahr2024l4acados}. With this process it is possible to learn a more accurate model for the MPC framework. Generally, to integrate a neural network into the MPC framework, the network's output is linearly added to the dynamics and is defined through the \textit{neural-enhanced} optimization problem
\begin{subequations} \label{eq:fused_problem}
    \begin{align}
        \underset{\boldsymbol{u}_k}{\rm min} \quad &\sum\limits_{k=0}^{N-1}\left(\overline{\boldsymbol{x}}\transpose_k \boldsymbol{Q} \overline{\boldsymbol{x}}_k + \overline{\boldsymbol{u}}\transpose_k \boldsymbol{R} \overline{\boldsymbol{u}}_k\right) + \overline{\boldsymbol{x}}\transpose_N \boldsymbol{Q}_N\overline{\boldsymbol{x}}_N, \label{eq:cost_fused} \\
        {\rm s.t.} \quad
        &\boldsymbol{x}_{k+1} = \underbrace{\boldsymbol{f}_{\mathcal{A}}\left(\boldsymbol{x}_{k}, \boldsymbol{u}_{k}\right) + \tilde{\boldsymbol{f}}\left(\boldsymbol{x}_{k}, \boldsymbol{u}_{k}\right)}_{=:\boldsymbol{f}_{\mathcal{N}}\left(\boldsymbol{x}_{k}, \boldsymbol{u}_{k}\right)}, \label{eq:dyn_constraint_fused} \nonumber \\
        &\quad\quad\quad\quad\quad\quad\quad\quad\quad\quad\quad\, k=0, \ldots, N-1, \\
        &\boldsymbol{x}_0 = \hat{\boldsymbol{x}}, \label{eq:init_constraint_fused} \\
        &\boldsymbol{x} \in \mathcal{X}, \boldsymbol{u} \in \mathcal{U} \label{eq:space_constraint_fused}
    \end{align}
\end{subequations}
with the analytical model $\boldsymbol{f}_{\mathcal{A}}\left(\boldsymbol{x}_{k}, \boldsymbol{u}_{k}\right)$ following the dynamics stated in section~\ref{sec:analytical_model}. The neural-enhanced model $\boldsymbol{f}_{\mathcal{N}}\left(\boldsymbol{x}_{k}, \boldsymbol{u}_{k}\right)$ is the sum of the analytical model and a fully connected neural network $\tilde{\boldsymbol{f}}\left(\boldsymbol{x}_{k}, \boldsymbol{u}_{k}\right)$.

Our approach for training the neural network to represent the residual dynamics is to generate suitable labels from recorded real-world data by comparing the measured state $\hat{\boldsymbol{x}}_{i+1}$ with the internally propagated state $\boldsymbol{x}_{\mathcal{A}_{i+1}}$ based on the initial measurement $\hat{\boldsymbol{x}}_{i}$. The actual state as well as the internal state is measured continuously for each step $i$ in $T_\mathrm{step} =$ \SI{0.1}{s} intervals. The internal state at step $i+1$ is computed by integrating the analytical model with a Runge-Kutta fourth-order integrator resulting in
\begin{equation}
    \boldsymbol{x}_{\mathcal{A}_{i+1}} = \mathrm{RK4}\left(\boldsymbol{f}_{\mathcal{A}}\left(\boldsymbol{\hat{x}}_{i}, \boldsymbol{u}^*_{0,i} \right)\right).
\end{equation}

The goal of the neural network $\tilde{\boldsymbol{f}}$ is to predict the residual between the real-world behavior and the behavior predicted by analytical model captured in $\boldsymbol{x}_{\mathcal{A}_{i+1}}$ but to simplify the training, we select the following dimensions of the state and control input as the input to the network
\begin{equation}
    \boldsymbol{\xi}_i = \left[ z_i, \boldsymbol{v}_i, \boldsymbol{q}_i, \boldsymbol{\alpha}_{s_i}, f_{r_i} \right], \quad r \in[1,... ,4]
\end{equation}
and we only focus on the residual linear acceleration $\tilde{\boldsymbol{a}}$ as the output of network. The residual acceleration $\tilde{\boldsymbol{a}}$ represents the main model error of the linear dynamics in~\eqref{eq:linear_analytical_dynamics}. Thus, we define the label $\boldsymbol{a}'_i$ that the neural network is trained to predict to be the residual acceleration
\begin{equation}
    \tilde{\boldsymbol{a}}_i \xrightarrow{} \boldsymbol{a}'_i =  \hat{\boldsymbol{a}}_i - \boldsymbol{a}_{\mathcal{A}_{i}}
\end{equation}
with $\boldsymbol{a}_{\mathcal{A}_{i}}$ as the acceleration component of the integrated state $\boldsymbol{x}_{\mathcal{A}_{i+1}}$. We numerically differentiate the acceleration to receive the closed form 
expressed in components of the state
\begin{equation}
    \tilde{\boldsymbol{a}}_i \xrightarrow{} \boldsymbol{a}'_i = \frac{\boldsymbol{\hat{v}}_{i+1} - \boldsymbol{\hat{v}}_{i}}{t_{i+1} - t_i} - \frac{\boldsymbol{v}_{\mathcal{A}_{i+1}} - \boldsymbol{\hat{v}}_{i}}{t_{i+1} - t_i} = \frac{\boldsymbol{\hat{v}}_{i+1} - \boldsymbol{v}_{\mathcal{A}_{i+1}}}{t_{i+1} -t_i},
\end{equation}where $\boldsymbol{v}_{\mathcal{A}_{i+1}}$ is the velocity component of the state propagation $\boldsymbol{x}_{\mathcal{A}_{i+1}}$.

\subsection{Energy-based Regularization} \label{sub:energy_based_regularization}
Our main contribution stems from designing a novel regularization technique for training the network to predict the label $\boldsymbol{a}'_i$ through energy considerations. When using supervised machine learning for neural networks the $\mathcal{L}_{2}$ norm is widely used as the loss function to train the network. Given the network prediction $\tilde{\boldsymbol{a}}_i$ and the label $\boldsymbol{a}'_i$, the $\mathcal{L}_{2}$ norm
\begin{equation}
    \mathcal{L}_{2}\left(\boldsymbol{a}'_i,  \tilde{\boldsymbol{a}}_i\right) = \left|\left|\boldsymbol{a}'_i- \tilde{\boldsymbol{a}}_i \right|\right|^2_2
\end{equation}
directly punishes predictions far away from the label.

We propose an additional regularization loss function $\mathcal{L}_E\left(\boldsymbol{\xi}_i,\tilde{\boldsymbol{a}}_i\right)$ based on the energy of the system given the current state-input vector $\boldsymbol{\xi}_i$ and predicted output $\tilde{\boldsymbol{a}}_i$. The idea is to penalize the neural network when it causes control corrections, that induce a large increase in the model-predicted system energy. The neural network trained under the $\mathcal{L}_{2}$ norm learns to match the labels as best as possible but it disregards physical feasibility. Our energy-based loss aims to balance the impact of the input on the output implicitly by measuring the energy change of the predicted state. By introducing this loss we train the network to encode the physical effect the network output has on the system. We define
\begin{equation}\label{eq:energy_loss}
    \mathcal{L}_E\left(\boldsymbol{\xi}_i, \tilde{\boldsymbol{a}}_i\right) :=  \mathrm{ReLU} \left(\Delta E\left(\boldsymbol{\xi}_i, \tilde{\boldsymbol{a}}_i\right) \right)
\end{equation}
as the \textit{energy-based regularization loss}, with the energy difference
\begin{align}\label{eq:energy_difference}
    \Delta E\left(\boldsymbol{\xi}_i, \tilde{\boldsymbol{a}}_i\right) = 
    &\underbrace{\frac{1}{2} m \left( \left|\left|\boldsymbol{v}_{\mathcal{N}_{i+1}}\left(\boldsymbol{\xi}_i, \tilde{\boldsymbol{a}}_i\right)\right|\right|_2^2 - \left|\left|\boldsymbol{v}_{\mathcal{A}_{i+1}}\left(\boldsymbol{\xi}_i\right)\right|\right|_2^2\right)}_{\mathrm{Kinetic\,Energy}} \nonumber \\
    & + \underbrace{\vphantom{\frac{1}{1}}m g \left(
    h_{\mathcal{N}_{i+1}}\left(\boldsymbol{\xi}_i, \tilde{\boldsymbol{a}}_i\right) - h_{\mathcal{A}_{i+1}}\left(\boldsymbol{\xi}_i\right)\right)}_{\mathrm{Potential\,Energy}}
\end{align}
where the velocities $\boldsymbol{v}_{\mathcal{N}_{i+1}}$ and $\boldsymbol{v}_{\mathcal{A}_{i+1}}$ as well as the heights $h_{\mathcal{N}_{i+1}}$ and $h_{\mathcal{A}_{i+1}}$ are components of the state vector after being propagated by the neural-enhanced dynamic $\boldsymbol{f}_{\mathcal{N}}$ and the analytical model $\boldsymbol{f}_{\mathcal{A}}$ respectively. Specifically, the state components result from integrating the analytical dynamics
\begin{align}
    h_{\mathcal{A}_{i+1}}\left(\boldsymbol{\xi}_i\right) &=
    \mathrm{RK4}_{\,h}\left( \boldsymbol{f}_{\mathcal{A}}\left(\boldsymbol{\xi}_i\right)\right) \\
    \boldsymbol{v}_{\mathcal{A}_{i+1}}\left(\boldsymbol{\xi}_i\right)
    &=
    \mathrm{RK4}_{\,\boldsymbol{v}}\left( \boldsymbol{f}_{\mathcal{A}}\left(\boldsymbol{\xi}_i\right)\right)
\end{align}
and the neural-enhanced dynamics
\begin{align}
    h_{\mathcal{N}_{i+1}}\left(\boldsymbol{\xi}_i, \tilde{\boldsymbol{a}}_i\right) &=
    \mathrm{RK4}_{\,h}\left( \boldsymbol{f}_{\mathcal{A}}\left(\boldsymbol{\xi}_i\right) + \tilde{\boldsymbol{a}}_i\right) \\
    \boldsymbol{v}_{\mathcal{N}_{i+1}}\left(\boldsymbol{\xi}_i, \tilde{\boldsymbol{a}}_i\right)
    &=
    \mathrm{RK4}_{\,\boldsymbol{v}}\left( \boldsymbol{f}_{\mathcal{A}}\left(\boldsymbol{\xi}_i\right) + \tilde{\boldsymbol{a}}_i\right).
\end{align}


With our proposed loss in~\eqref{eq:energy_loss} we capture the energy of the state $\boldsymbol{x}_{\mathcal{N}_{i+1}}$ that is reached with the network output. We compare this to the energy of the state $\boldsymbol{x}_{\mathcal{A}_{i+1}}$ that is predicted by the analytical model. We define the loss function as the Rectified Linear Unit (ReLU) precisely to penalize the network if the output generates a higher energy state but not take effect if it results in a lower energy state.

Finally, the total loss function combines the standard $\mathcal{L}_{2}$-loss and our energy-based regularization to
\begin{equation}
    \mathcal{L}\left(\boldsymbol{a}', \tilde{\boldsymbol{a}}\right) = \mathcal{L}_{2}\left(\boldsymbol{a}', \tilde{\boldsymbol{a}}\right) + \lambda_E \mathcal{L}_E\left(\boldsymbol{a}', \tilde{\boldsymbol{a}}\right)
\end{equation}
with the empirically derived weight $\lambda_E = 10^3$.

\section{Experiments}\label{sec:experiments}
\begin{figure}
    \centering
    \includegraphics[width=\linewidth]{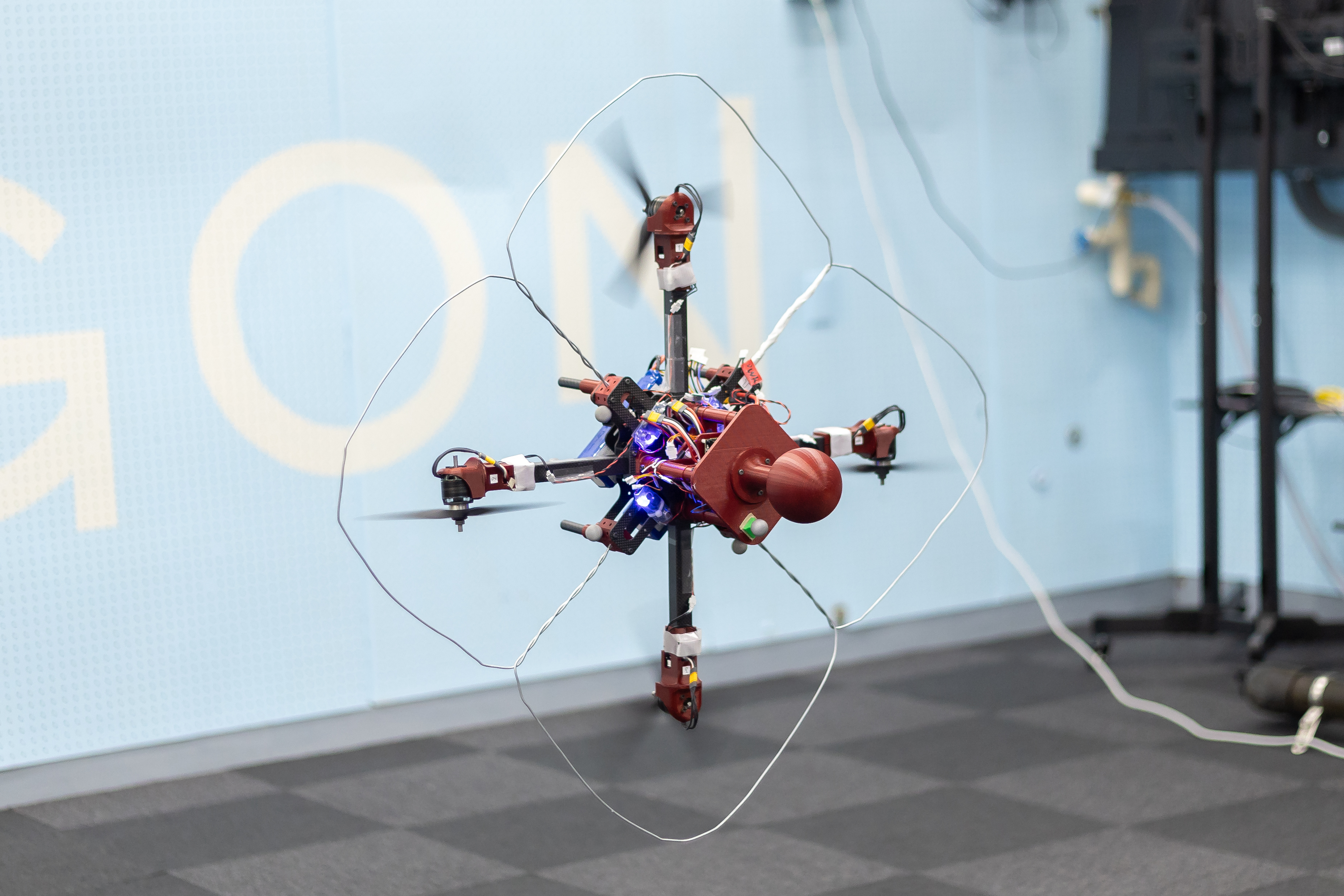}
    \caption{Pose during a data gathering experiment where the omnidirectional aerial robot performs various rotation trajectories such as \ang{90} (as seen above) and even \ang{180} turns. The robot achieves omnidirectional flight through servo motors at the end of each arm that actuate the angle of attack of each thrust vector w.r.t.\ the body frame.}
    \label{fig:robot}
\end{figure}
For training neural networks in our work we initially record flight data controlling the quadrotor using the analytical model, i.e., using $\tilde{\boldsymbol{f}}\left(\boldsymbol{x}_{k}, \boldsymbol{u}_{k}\right) \equiv \boldsymbol{0}$. We track the measured state $\hat{\boldsymbol{x}}_i$ and optimized input $\boldsymbol{u}^*_{0,i}$ at each time step $i$. An OptiTrack Motion Capture system is used to externally observe the pose of the quadrotor in \SI{100}{Hz} frequency. With this method we can obtain accurate measurements for the position $\hat{\boldsymbol{p}}_i$ and attitude $\hat{\boldsymbol{q}}_i$.
An Extended-Kalman-Filter (EKF) is used to fuse measurements from other sensors to build a full state estimate $\boldsymbol{\hat{x}}_i$. The omnidirectional aerial robot is depicted during a roll-rotation experiment mid-flight in Figure~\ref{fig:robot}. For the MPC framework we use the \texttt{acados}~\cite{verschueren2022acados} framework together with the partial condensing \texttt{HPIPM}~\cite{frison2020hpipm} as SQP real-time iteration solver for all controllers.


\begin{table}[t]
  \centering
  \caption{Positional MAE for each experiment using three controllers.}
  \label{tab:tracking_errors}

  \begin{adjustbox}{max width=\columnwidth}
    \setlength{\tabcolsep}{6pt}
    \renewcommand{\arraystretch}{1.10}
    \begin{tabular}{l S[table-format=1.4] S[table-format=1.4] S[table-format=1.4]}
      \toprule
      {Controller} & {Setpoint [m]} & {Circle [m]} \\
      \midrule
      Analytical MPC & 0.1283 & 0.1593 \\
      RTNMPC         & \bfseries 0.0806 & 0.1268 \\
      Ours           & 0.0816 & \bfseries 0.1072 \\
      \bottomrule
    \end{tabular}
  \end{adjustbox}
\end{table}

The framework used to train neural networks in this work is \texttt{PyTorch}~\cite{Paszke2019PyTorchAI} with the AdamW optimizer~\cite{loshchilov2017decoupled}, a batch size of 64 over 130 epochs and weight decay of $10^{-3}$. We choose the learning rate $\mathrm{lr}(n) = \max\left(10^{-3} \cdot 0.99^{n}, \, 10^{-5}\right)$ as a function of the epoch $n$. The network is a classical Multi-Layer-Perceptron (MLP) with a single hidden layer containing 32 nodes applying a GELU~\cite{hendrycks2016gaussian} activation function.
We recorded 15 minutes of flight including maneuvers close to the ground as well as tilt and roll trajectories.

\begin{figure}[t!]
    \centering
    \includegraphics[width=\linewidth]{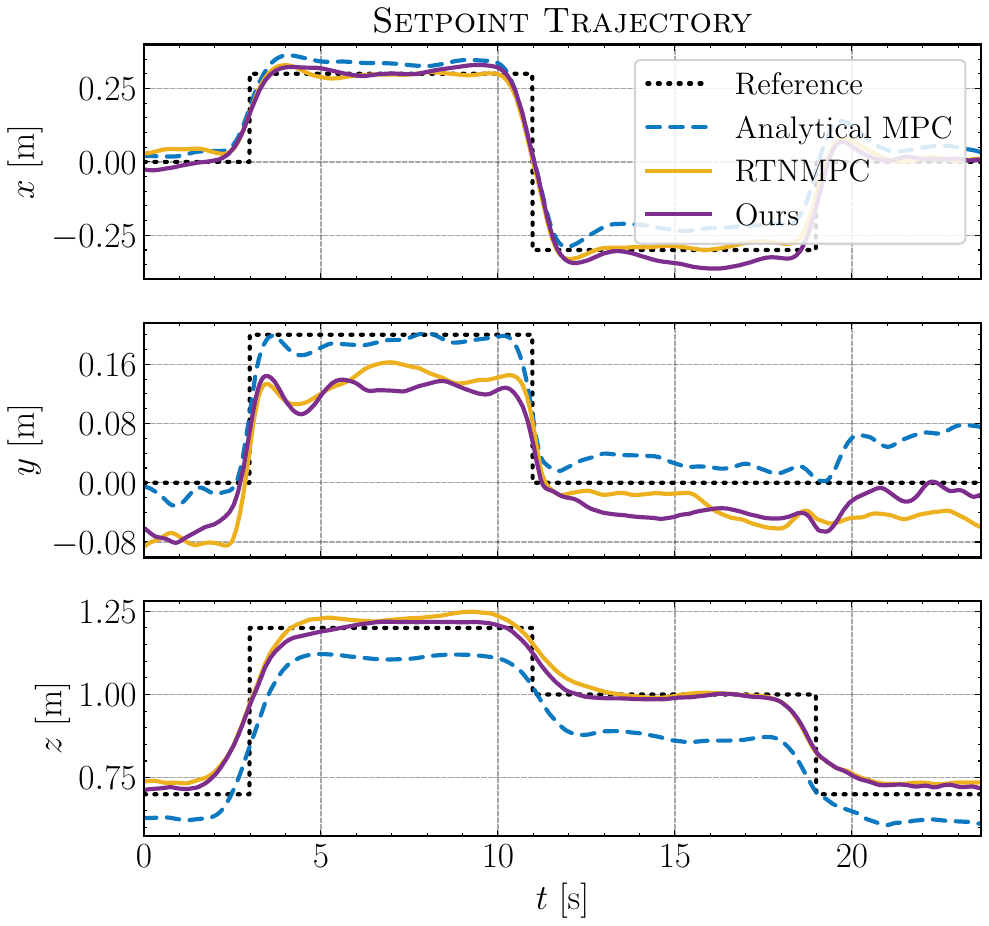}
    \caption{Comparison of the position $\left[ x, y, z \right]$ following a setpoint tracking scenario while using either a analytical MPC, a standard neural MPC (RTNMPC) or our proposed method; it shows that both neural controllers can compensate model errors and match the reference well in $\mathbf{x}$ and $\mathbf{z}$ but perform worse than the analytical MPC in $\mathbf{y}$.}
    \label{fig:position_setpoint}
\end{figure}

\begin{figure}[h!]
    \centering
    \includegraphics[width=\linewidth]{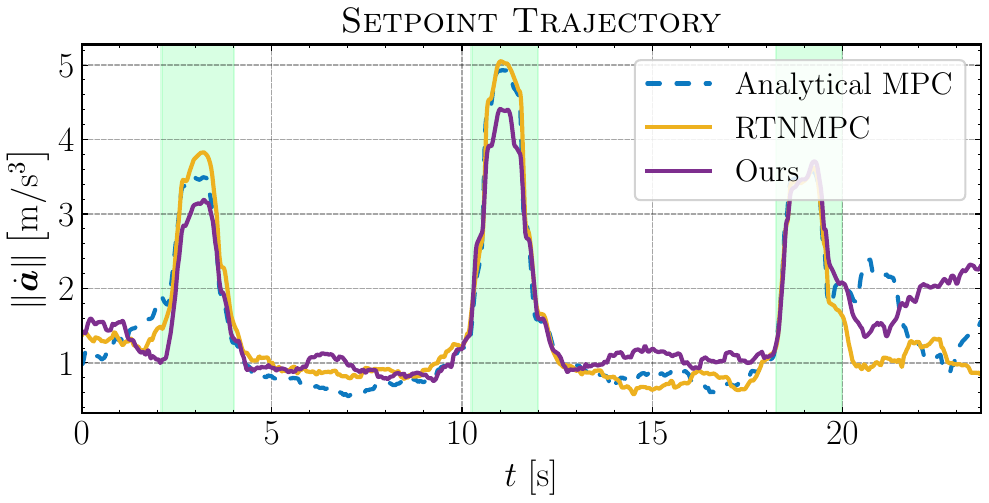}
    \caption{Comparison of the jerk norm $\|\dot{\boldsymbol{a}}\|$ while tracking challenging setpoints using either an analytical MPC, a standard neural MPC (RTNMPC) or our proposed method. The jerk is differentiated from the acceleration after applying a low-pass filter at \SI{10}{Hz}. It shows that the acceleration change is significantly stabler for our method which is a direct consequence of training the network under the proposed energy-based regularization loss.}
    \label{fig:smoothness_setpoint_only}
\end{figure}

To validate our proposed method we conducted two experiments, firstly tracking two setpoints with challenging poses and secondly following a circle trajectory. Each experiment compares three different controllers: analytical MPC, standard neural MPC and our energy-based neural MPC. As a representative for many other methods that build on the standard neural MPC formulation but do not use any additional regularization techniques we benchmark our work to Real-Time Neural MPC (RTNMPC) from Salzmann et al.~\cite{salzmann2023real}. To allow for accurate comparison the only difference from our method to our implementation of RTNMPC is the energy-based regularization loss applied during training. The MAE for the position $\overline{\boldsymbol{p}}$ are given for each trajectory scenario in Table~\ref{tab:tracking_errors}.

\subsection{Setpoint Trajectory}
The first experiment is conducted by tracking two poses requiring the omnidirectionality of the tiltable-quadrotor platform. Figure~\ref{fig:position_setpoint} shows the position $\left[ x, y, z \right]$ while tracking the poses with each controller. Here it can be seen that both neural MPCs are capable of reducing the steady-state error in the $\mathbf{z}$-axis. The larger error in the $\mathbf{y}$-axis is small in absolute measure. However, in total making use of the neural model improves the tracking performance. For transparency, the performance of our method shown in Table~\ref{tab:tracking_errors} is worse for this scenario than RTNMPC which can be explained from data noise. Furthermore, the neural model doesn't compensate the angular acceleration nor does the angular moment influence the energy cost. These will be vantage points for further research and are prognosed to further increase accuracy. In this scenario we primarily show the compensation of model error while the robot is tracking static targets but during hovering our method cannot differ significantly to the baseline RTNMPC resulting in similar overall tracking performance. Only in the transition movements between setpoints we can observe that the energy loss is taking effect. Figure~\ref{fig:smoothness_setpoint_only} shows the jerk norm for this experiment. Since the controller can only affect the robot's motion at low frequencies, we smoothened the acceleration data with a low-pass filter at \SI{10}{Hz} before differentiating. It can be seen that especially in the highlighted transition areas our model exerts significantly lower jerk. Therefore, the acceleration changes for our proposed method in a more stable fashion than the acceleration using RTNMPC or the analytical MPC. Since the kinetic energy is directly correlated with the acceleration, this showcases that the introduction of penalizing the network with the proposed energy-based regularization~\eqref{eq:energy_loss} results energy-conserving flight. It further supports our claim that the network is able to minimize the physical impact on the system's energy. The neural model proves to learn the model error while stabilizing the predicted acceleration and therefore enabling a smoother flight.

\subsection{Circle Trajectory}
\begin{figure}[t!]
    \centering
    \includegraphics[width=.918\linewidth]{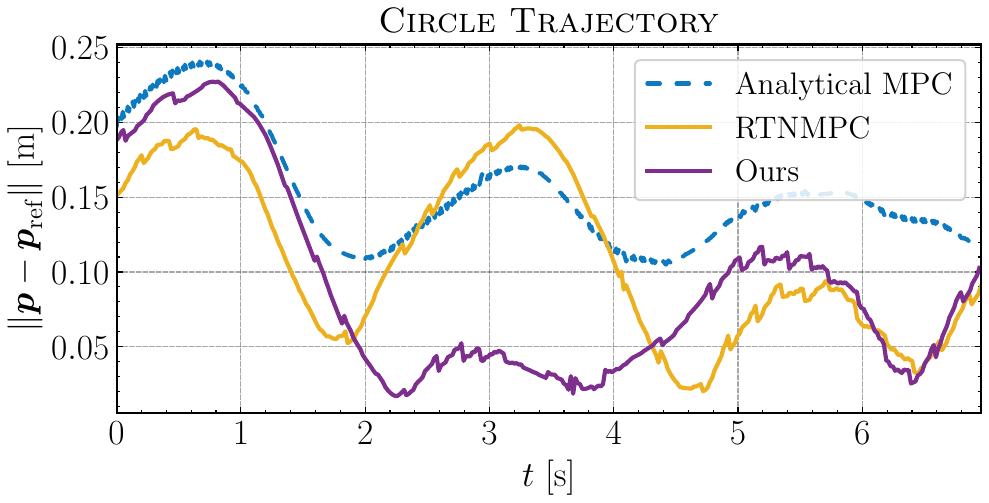}
    \caption{Comparison of the positional tracking error $\overline{\boldsymbol{p}}$ following a planar circle scenario while using either a analytical MPC, a standard neural MPC (RTNMPC) or our proposed method; it shows that our method manages to find control inputs that have the smallest MAE due minimizing the relative energy.}
    \label{fig:tracking_error_circle}
\end{figure}

For our second experiment we navigate the robot in a circle trajectory in the $\mathbf{x}$-$\mathbf{y}$-plane. The tracking error is given in Figure~\ref{fig:tracking_error_circle}. This experiment showcases the advantage of our proposed method compared to the standard neural MPC. Our method manages to find predictions that induce low-acceleration movement which in turn allows the robot to track the reference better than not only the analytical MPC but also RTNMPC. This is achieved due to the neural network outputting physically accurate corrections to the model enabling an energetically stable flight compared to the naive network outputs in the standard neural MPC. The network successfully learned to encode the physical effect its output has on the system's energy and therefore allows the MPC to predict the acceleration in a stabilized manner resulting in a more accurate trajectory.


\section{Conclusion} \label{sec:conclusion}
In this work we introduce a novel \textit{energy-based regularization} for data-driven MPC. The regularization is applied to the training of a neural network which learns the residual dynamics of an omnidirectional aerial robot. Our proposed loss penalizes the network it its output generates a higher energy in the system than only using the analytically modeled system. Due to this regularization the network is encouraged to predict outputs that have lower acceleration changes and are therefore more stable. This allows the MPC framework to optimize suitable control inputs. We show that this approach outperforms not only the analytical but also a standard neural controller in trajectory tracking. We achieve a positional MAE of \SI{0.1122}{m} compared to \SI{0.1466}{m} for the analytical MPC averaged over three scenarios in real-world experiments. Our approach also improves the MAE by up to 15\% compared to the standard neural MPC in selected experiments. Notably, the focus on the system's relative energy during training improves not only the flight stability but also the flight accuracy. For future work, we postulate that using more advanced network architectures as well as integrating the angular acceleration into the framework can leverage the energy-based regularization further.




\balance
\bibliographystyle{Bibliography/IEEEtran}
\bibliography{Bibliography/IEEEabrv, Bibliography/references}

\end{document}